# Investigation of photon emitters in Ce-implanted hexagonal boron nitride

Gabriel I. López-Morales,[a,b,c] Mingxing Li,[a] Alexander Hampel,[d] Sitakanta Satapathy,[a] Nicholas V. Proscia,[a] Harishankar Jayakumar,[a] Artur Lozovoi,[a] Daniela Pagliero,[a] Gustavo E. Lopez,[b,c] Vinod M. Menon,[a,c] Johannes Flick,[d] Carlos A. Meriles[*,a,c]

[a] *Department of Physics, City College of the City University of New York, New York, NY 10031 USA*
[b] *Department of Chemistry, Lehman College of the City University of New York, Bronx, NY 10468, USA*
[c] *The Graduate Center of the City University of New York, New York, NY 10016, USA*
[d] *Center for Computational Quantum Physics, Flatiron Institute, New York, NY 10010, USA*
[*] *cmeriles@ccny.cuny.edu*

**Abstract:** Color centers in hexagonal boron nitride (hBN) are presently attracting broad interest as a novel platform for nanoscale sensing and quantum information processing. Unfortunately, their atomic structures remain largely elusive and only a small percentage of the emitters studied thus far has the properties required to serve as optically addressable spin qubits. Here, we use confocal fluorescence microscopy at variable temperature to study a new class of point defects produced via cerium ion implantation in thin hBN flakes. We find that, to a significant fraction, emitters show bright room-temperature emission, and good optical stability suggesting the formation of Ce-based point defects. Using density functional theory (DFT) we calculate the emission properties of candidate emitters, and single out the $CeV_B$ center — formed by an interlayer Ce atom adjacent to a boron vacancy — as one possible microscopic model. Our results suggest an intriguing route to defect engineering that simultaneously exploits the singular properties of rare-earth ions and the versatility of two-dimensional material hosts.

Solid-state quantum registers formed by interacting electron and nuclear spins amenable to high-fidelity state manipulation and readout provide a promising architecture for quantum technologies, including local information processing and storage, error correction, and long-distance state transmission via photon emission [1,2]. Pioneering work during the last decade has propelled color centers in diamond, silicon carbide, and other wide-bandgap semiconductors as reference platforms in the search for optimal chip-integrated solid-state quantum processors [3–5]. Adding to this list, recent effort has been devoted to the investigation of photon emitters in two-dimensional (2D) materials [6,7], where ever-improving 'pick-and-place' techniques promise opportunities for integration with electronic or photonic structures [8–10].

From among a growing set of options, hexagonal boron nitride (hBN) is attracting broad interest as the single wide-bandgap van der Waals host of color centers with narrow emission linewidths [11,12], high spectral-tunability [13,14], and strain activation [15]. Moreover, recent work showed the emission from some of these defects exhibits a magneto-optical response [16], ultimately exploited to demonstrate optically-detected magnetic resonance (ODMR) [17,18]. Unfortunately, it is yet unclear whether these intrinsic defects — whose exact physical nature is only now being uncovered [19] — can be more controllably engineered, and whether the strong hyperfine interactions with surrounding nuclei — all of which are spin-active — are compatible with efficient spin-qubit manipulation protocols.

Here we explore an alternative family of defects in hBN derived from low-energy implantation of rare earth ions. The latter are attractive in that their partly filled 4f electron orbitals are screened from the outside by outer-lying electrons, which makes their response less sensitive to the host crystal. Further, 4f electrons are often unpaired and hence exhibit magnetic moments that can be polarized and probed. For the present work, we focus on Ce, a rare-earth atom studied in other wide bandgap hosts down to the level of single emitters [20]. Cerium has already been incorporated into hBN as a way to modify the crystal electronic structure and photoluminescence (PL) properties [21–24], although the doping concentrations ($10^2$-$10^3$ ppm) are greater than those explored herein (2–6 ppm).

We study Ce-implanted flakes of hBN on a $SiO_2$ substrate featuring an array of 40-μm-deep pits. From white-light optical measurements [25,26], we determine the thickness to range from ~50 to 100 nm, consistent with prior observations relying on the same dry-transfer protocols [27–29]; Raman microscopy measurements indicate the crystal quality is high [30–32] (see Supplemental Material, Section S1 for details). Room-temperature confocal microscopy reveals patchy areas of PL equally scattered across supported and suspended hBN regions (dashed white square in Fig. 1a), hence pointing to in-flake emitters minimally impacted by the substrate. Since flakes are virtually emitter-free prior to implantation, we conclude the observed color centers stem from Ce bombardment. This process, however, is likely to produce a broad class of defects hosting (or not) a Ce ion, a notion our experimental results seem to confirm.

Specifically, Fig. 1b shows the spectral PL response from representative sites across the flake, exhibiting either broad emission (Type I), or narrow fluorescence lines (Type II); 'mixed' spectra — where both types coexist — are also common. Further, Type I emitters are found to be remarkably photostable under continuous excitation, while Type II are found to blink persistently and ultimately bleach after prolonged laser exposure. While unambiguous assignment is difficult, we associate Type II spectra with intrinsic hBN color centers created during ion implantation, likely vacancy- or antisite-based complexes analogous to those obtained from electron irradiation and bombardment with other species, such as carbon or oxygen [19,33,34]. Type I spectra, on the other hand, are more intriguing as they seem to originate from a different class of color center we tentatively associate to Ce-based defects.

To interpret our observations, we begin with the Dieke-Carnall energy diagram of $Ce^{3+}$, a characteristic charge state used here as an initial guide (Fig. 1c) [35]. With only one electron in its 4f orbital, the $^2F$ ground state breaks into a manifold of seven Kramers doublets. Since the separation between them is relatively large, selective excitation from (to) the lowest energy doublet in the 4f (5d) set is possible with a laser tuned at (or near) 489 nm (blue arrow in Fig. 1c) [20,35]. Note that because 5d levels correspond to outer lying orbitals, excited state lifetimes in $Ce^{3+}$ are expected to be short. Therefore, unlike heavier rare-earth ions — featuring low-photon-yield, intra-4f transitions — $Ce^{3+}$ defects are expected to be bright.

Admittedly, however, the connection between $Ce^{3+}$ ions and Type I emission is not immediate, as Type I spectra do not display a visible zero-phonon line (ZPL) near 489 nm, and their peak fluorescence wavelengths are heterogeneously distributed over a broad range. On the other hand, both observations are not unexpected as the Huang-Rhys factor of $Ce^{3+}$ defects in common hosts is high (due to electronic-confinement mismatch following 4f↔5d atomic transitions) [20,35,36], whereas strain- and/or local-charge-induced spectral shifts (also affecting Type II emitters, Fig. 1d) are known to be large [13,34,37].

Despite the present limitations, valuable clues emerge from a more extensive optical characterization. Fig. 2 zeroes in on a group of Type I sites, A through C, with nearly identical emission spectra and comparable excited state lifetimes. We monitor the fluorescence from one of these sites (A in Fig. 2a) as we vary the excitation wavelength using a tunable laser (Fig. 2b). Except for an overall scaling, we find the emission spectrum remains virtually unchanged, gradually fading as the excitation wavelength exceeds ∼ 490 nm. This response is consistent with that anticipated for a (hypothetical) $Ce^{3+}$-based defect, as anti-Stokes-assisted photon absorption must decrease below the $4f \leftrightarrow 5d$ transition energy [38]. Note that presumably identical sites (e.g., B, C in Fig. 2a), show different photoluminescence excitation (PLE) spectra (Fig. 2c), thus hinting at several subclasses of Type I emitters (the most likely scenario following an implantation process).

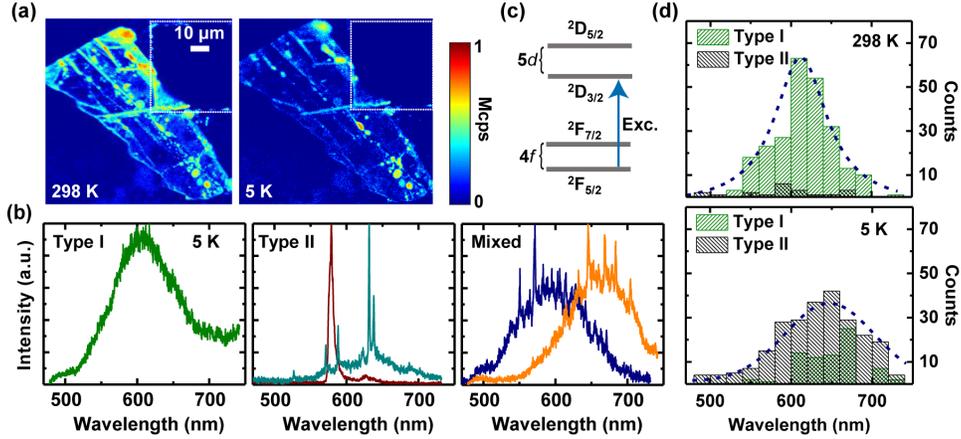

Fig. 1. Steady-state spectroscopy of Ce-implanted hBN flakes. (a) Confocal microscopy images of a Ce-implanted hBN flake under 460 nm laser excitation at room-temperature (left) and 5 K (right) revealing a high density of emitting sites across the flake. Emission stemming from a suspended region of the flake (dashed white square) indicates emitters are localized within hBN, not at the interface with the substrate. (b) Representative PL spectra collected from different sites throughout the hBN flake, revealing the simultaneous presence of Ce-derived defects (Type I) and intrinsic hBN defects (Type II). Noticeable spectral shifts are often observed for both types of defects. (c) Energy-level diagram for $Ce^{3+}$. (d) Statistics of Type I and Type II spectra collected from isolated sites across multiple hBN flakes at room-temperature (upper panel) and 5 K (lower panel). Dashed lines are guides to the eye.

Although these results suggest the 'atom-like' properties of Ce ions are somewhat preserved, the very notion of a Ce-based defect in hBN is a priori unclear, especially given the large atomic number mismatch between the implanted ions and the atoms of the crystal host. To gain some understanding, we resort to density-functional theory (DFT) in the generalized-gradient approximation (GGA) including corrections for van der Waals interactions [39–47]. Throughout our calculations, we use the projector augmented wave (PAW) method as implemented in the Vienna Ab-initio Simulation Package [48], with an orthorhombic supercell [49,50] of 6 layers, each containing 112 atoms (Supplemental Material, Section S.II.a).

Fig. 3 summarizes the DFT results from one of the considered defects, where cerium associates with a boron vacancy in one of the adjacent atomic planes (Fig. 3a). In the following, we denote this system as $CeV_B$ instead of $Ce_B$ due to the inter-layer position of the Ce atom. Unlike the case with a nitrogen vacancy, $CeV_N$ (Supplemental Material, Section S.II.b), we find reduced deformation of the atomic planes surrounding the defect. Calculation of the charge stability transition energies point to the positive, neutral, and negative charge states as the three most likely, with one or the other becoming dominant depending on the system's Fermi energy (Fig. 3b). The latter can be modified locally via strain, which, in turn, can lead to pockets of fluorescent emitters heterogeneously distributed across the hBN flake; we suspect this mechanism underlies the formation of the bright spots seen in Figs. 1a and 2a.

Inspection of the calculated density-of-states (DOS) plots in Fig. 3c-e reveals intra-bandgap levels deriving from the CeV$_B$ defect. The neutral charge state DOS (Fig. 4c) shows a single 4*f* state in the bandgap as the highest occupied orbital. Below the 4*f* state, other localized defect sates — corresponding to the boron vacancy — also fall inside the bandgap. Similarly, the negative charge state in Fig. 3d shows an occupation of a 4*f* state as well as an orbital with significant 5*d* character close to the conduction band. In addition, we find low hybridization of the ion's 4*f* and 5*d* levels with localized defect states from the boron vacancy or bulk hBN, suggesting this system retains some of the key properties of the bare Ce$^{3+}$ ion. We emphasize this feature cannot be generalized to all Ce-related defects as we find strong hybridization, e.g., when the Ce atom is forced to take an in-plane position (Supplementary Material, Section S.II.c).

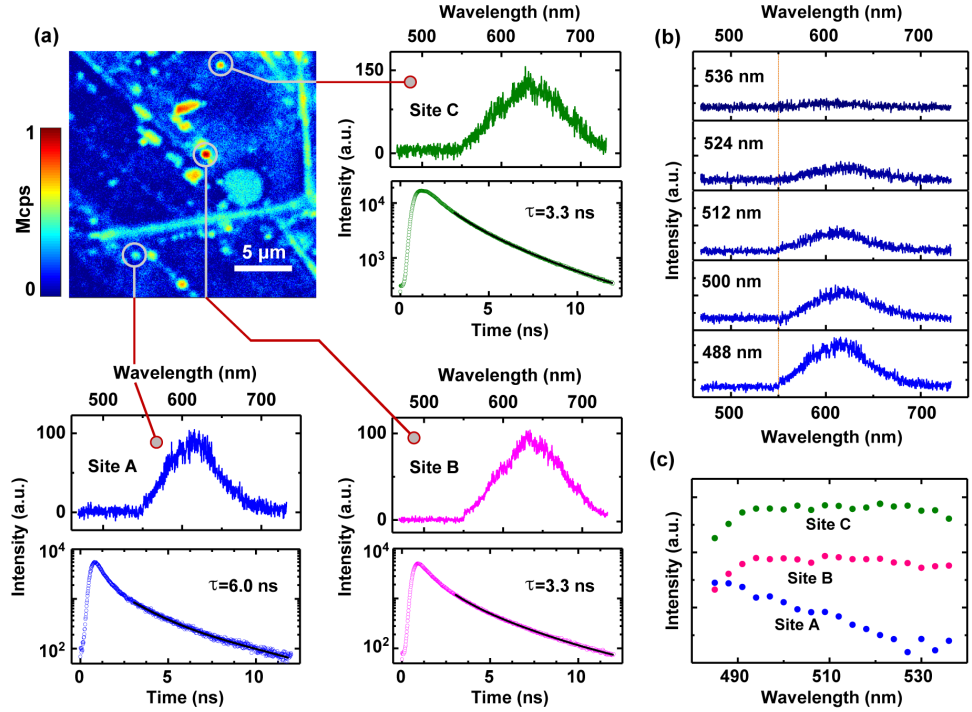

Fig. 2. Characterization of Type I defects. (a) (Main) Confocal image of a zoomed-in region from Figure 1(a) under 485 nm pulsed-laser excitation. (Side Inserts) Fluorescence spectra and time decay for from three representative sites across the flake. Single-exponential fits reveal decay constants within 3–10 ns. (b) PLE spectra for all three sites in (a); dashed vertical line is the cutoff wavelength for PL collection. (c) PLE integrated intensity. Sites B and C follow a different trend, hinting at different classes of defects.

We access excited state properties for this system by resorting to the self-consistent field (ΔSCF) approach (Fig. 3f). ΔSCF considers the energy difference between two separate energy calculations, one for the ground state and a second one, where we promote an electron from an occupied 4*f* state to a higher energy 5*d*-orbital. We follow the ab-initio methods outlined in [51] (Eqs. 2–11), which have been applied to NV$^-$ centers in diamond and other atomic defects in hBN [51–53]. These methods are based on a generating function and make use of the coupling between the ionically relaxed ground and excited state configurations, allowing to calculate optical properties (such as the Huang-Rhys factors). Figs. 3e-g present the excited state properties for the defect structure in Fig. 3a, obtained within this approach. We find a large configurational mismatch between the excited and ground state orbitals leading to a Huang-Rhys factor of 27 and a broad PL spectrum with vanishing ZPL (Fig. 3g), in qualitative agreement with the results summarized in Figs. 1 and 2. Note, however, that other potential Ce-vacancy complexes could show similarly broad features.

In summary, Ce implantation of hBN flakes produces a broad class of photon emitters,

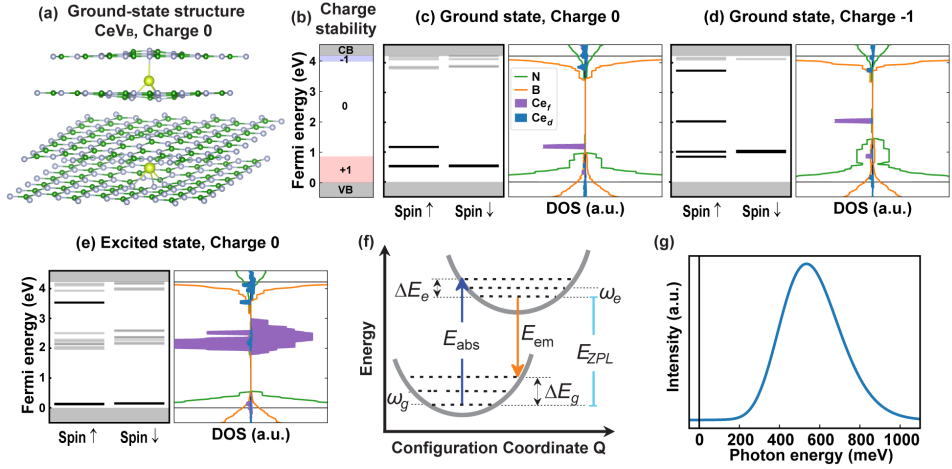

Fig. 3. Theoretical modelling of the $CeV_B$ defect center in hBN. (a) Atomic structure of the $CeV_B$ center (green, yellow, and gray spheres represent B, Ce and N atoms, respectively). (b) Charge stability for the +1, 0 and -1 charge states. (c-d) Ground state energy levels and density of states (DOS) for the 0 and -1 charge states, respectively. (e) Excited state energy levels and DOS for the neutral (0) charge state only. (f) The general 1D configurational-coordinate diagram, constructed from ground and excited state electronic calculations. (g) The corresponding emission spectrum as calculated from the effective parameters identified in (f). We calculate a negligible zero-phonon emission (vertical line at the origin) and a broad Stokes-shifted spectrum, in qualitative agreement with our observations. For the ZPL — set to 0 meV in the figure — we find a value of 2.50 eV.

which we phenomenologically group in two classes featuring qualitatively different fluorescence spectra. One of these — exhibiting bright, stable Stokes-shifted emission over a broad band — generically reproduces $Ce^{3+}$ emission previously seen in other material hosts, hence suggesting the presence of the ion. PLE data indicates that not all emitters in this class behave identically; in particular, we identify one sub-class whose response as a function of the illumination wavelength matches that expected for a $Ce^{3+}$ ion. DFT calculations support the notion of stable Ce-based point defects in hBN. From among alternative options, the $CeV_B$ — comprising a Ce atom displaced from a vacant boron site towards an inter-layer position — features an intra-bandgap ground (excited) state with strong $4f$ ($5d$) character. Further, $\Delta$SCF calculations allow us to derive a broad emission spectrum with vanishing ZPL, in qualitative agreement with our observations.

More generally, our results suggest an intriguing route to emitter engineering that benefits from the singular properties of 2D materials without necessarily relying on intrinsic point defects. Indeed, the atom-like properties of rare-earth ions prompts us to think in terms of an elemental toolkit containing nearly-ready-to-use spin qubits. The latter appears relevant to systems such as hBN, where all nuclear isotopes are spin active and intrinsic magneto-optically-active defects exhibit broad ODMR lines [17,18] (at odds with spin control protocols). It is still unclear, however, whether color centers stemming from rare-earth ion implantation exhibit valuable spin properties, a question that will require further investigation. On the other hand, the use of multi-layer membranes makes it possible to create surface-insensitive environments, while facilitating integration with structures for light extraction, strain engineering, or voltage control. Finally, extending this approach to 2D systems with heavier host atoms — such as transition metal dichalcogenides — could prove possible despite their lower bandgap, e.g., by resorting to rare-earth ions such as $Er^{3+}$ with emission in the telecom band [54].

**Acknowledgements**

G.I.L.M. and N.V.P. acknowledge financial support from CREST IDEALS supported through grant NSF-1547830. V.M.M. and C.A.M. acknowledge support from the National Science Foundation through grant NSF-1906096. C.A.M. acknowledges support from Research Corporation through a FRED award. The Flatiron Institute is a division of the Simons Foundation.

## Disclosures

The authors declare no conflicts of interest.

## Data availability

The data that support the findings of this study are available from the corresponding author upon reasonable request.

## Supplemental document

See Supplemental Material for supporting content.